\documentclass[10pt,a4paper]{article}
\usepackage{graphicx}

\begin{document}
\textwidth=135mm
 \textheight=200mm

\begin{center}
{\bfseries Realistic and effective interactions in the study of nuclear matter
\footnote{{\small Talk given at the Helmholtz International Summer School
``Dense Matter in Heavy Ion Collisions and Astrophysics'', JINR, Dubna,
August 21 - September 1, 2006.}}}

\vskip 5mm

V. Som\`a 
\vskip 5mm

{\small {\it Institute of Nuclear Physics, PL-31-342 Cracow, Poland}}
\\

\end{center}

\vskip 5mm

\centerline{\bf Abstract} 
The equation of state of symmetric nuclear matter is addressed starting both 
from a realistic interaction derived from nucleon-nucleon scattering processes
and from a low-momentum effective potential. The approach is based on finite
temperature Green's functions. The internal energy per particle is estimated
from the summation of diagrams and through the Galitskii-Koltun's sum rule.
\vskip 10mm

At present, there is not a unique and reliable theory of nuclear interactions.
From the study of nucleon-nucleon scattering processes different models for 
the basic interaction between nucleons have been developed, but 
difficulties arise when they are used in calculations which aim to
reproduce the properties of bound nuclei.
These models all reproduce the experimental N-N phase shifts up to the
threshold for pion production with high accuracy, but differ in the
short-wavelength region, where less experimental constraints are
available.

In deriving the properties of dense nuclear matter, if we start from a bare 
N-N potential we have to take into account the complex correlations induced by
strong interactions.
In parallel with the development of diagram summation techniques, a variety
of effective potentials have been proposed. Apart from the traditional
phenomenological N-N forces, to be used in a mean-field approximation, such 
as the Skyrme and the Gogny interactions, in the last years renormalized
low-momentum potentials have been introduced \cite{rgfermi}. 
They are restricted to a subspace of the Hilbert space, the so-called model
space, in which they incorporate the short-range correlations between nucleons.
If the cutoff is sufficiently low, they turn to be independent of the
starting N-N interaction.

In this work we present a self-consistent scheme based on finite temperature
Green's functions and apply it to study the equation of state of
symmetric nuclear matter. We first consider a bare nucleon-nucleon 
interaction, namely the CD-Bonn potential,
and compare the results with the case in which a renormalized potential $V_{low-k}$
\cite{bodemu} is used.
\\

It is possible to construct a consistent approximation
starting from a suitably chosen generating functional
\cite{baym}. Such approximation
schemes automatically fulfill thermodynamic
relations, including the 
Hugenholz-Van Hove  and Luttinger identities.
For nuclear interactions the generating functional must at least include
ladder-type diagrams: this choice leads to the in-medium $T$-matrix,
the approximation scheme adopted in our study.

The in-medium two-particle scattering 
matrix $T$ is defined as (for simplicity we skip spin and isospin 
indices, as well as energy and momentum dependence):
\begin{equation}
\label{tmatrix}
T=V+V\:G_2^{nc} \: T \: \: .
\end{equation}
Here $V$ is the interaction potential and $G_2^{nc}$ is the 
\textit{non correlated} two-particle Green's function,
just constructed as a product of two dressed single-particle propagators.

The $T$-matrix accounts for the multiple scattering between nucleons and
thus permits to resum the short range correlations induced in the dense
system. We remark that
the use of the dressed propagator implies the presence of a nontrivial
dispersive self-energy which leads to a broad spectral function
\cite{Bozek:2002em}. 
The full two-particle propagator is then approximated as follows
\begin{equation}
\label{app}
\mathcal{G}_{2}=G_2^{nc}+G_2^{nc} \: T \: G_2^{nc}
+ \: exchange \: terms \: \: .
\end{equation}
Eqs. (\ref{tmatrix}) and (\ref{app}), together with the Dyson equation, 
form a scheme in which all ingredients
are calculated iteratively: the scattering matrix, the
single-particle self-energy, which is expressed in terms of
the scattering matrix, and the single-particle propagator.

Most of the properties of the system can be derived from these quantities.
The (total) internal  energy per particle can be calculated as the expectation
value of the Hamiltonian $H = H_{kin} + H_{pot}$. The two terms read
(we now make explicit the momentum and energy dependences)
\begin{equation}
\label{eq:kin}
\langle H_{kin} \rangle = \mathcal{V} \int \frac{d^3{p}}{(2\pi)^3} 
\frac{d \omega}{2\pi}
\frac{\mathbf{p}^2}{2m} A(\mathbf{p},\omega) f(\omega) \: ,
\end{equation}
where $\mathcal{V}$ represents the volume,
$A(\mathbf{p},\omega)$ is the spectral function and
$f(\omega)$ the Fermi-Dirac distribution, and
\begin{eqnarray}
\label{eq:final}
& \displaystyle
\langle H_{pot} \rangle = \frac{\mathcal{V}}{2} 
\int \frac{d^3{P}}{(2\pi)^3} \frac{d^3{k}}{(2\pi)^3} 
\frac{d\Omega}{2\pi}  \, b(\Omega) \, 
\nonumber \\
& \displaystyle \times
\mbox{Im} \left\{ \left( \langle \mathbf{k}|T^R(\mathbf{P},\Omega)
|\mathbf{k}\rangle-
 \langle \mathbf{k}|T^R(\mathbf{P},\Omega) |\mathbf{-k}\rangle\right)\,
{G_2^{nc\ R}}(\mathbf{P},\mathbf{k},\Omega)\right\} \: ,
\end{eqnarray}
where $\Omega$ and $\mathbf{P}$ are the total energy and momentum of the pair
of nucleons, $\mathbf{k}$ the exchanged momenta and
$b(\Omega)$ the Bose-Einstein distribution. 

An alternative and simpler way to determine the energy is the 
Galitskii-Koltun's sum rule \cite{galitskii,koltun}
\begin{equation}
\label{gksumrule}
\frac{E}{N} = \frac{1}{\rho}
\int \frac{d^3{p}}{(2 \pi)^3} \frac{d \omega}{2 \pi}
\left [ \frac{\mathbf{p}^2}{2m} + \omega \right ] 
A(\mathbf{p},\omega) f(\omega ) \: \: ,
\end{equation}
which for conserving approximations is 
equivalent to the direct calculation (\ref{eq:kin})+(\ref{eq:final}).
In the presence of three-body forces, however, this is not valid and the
expectation value of the Hamiltonian has to be computed to
estimate the internal energy of the correlated system.

The ladder expansion can be as well derived from a generating
functional $\Phi[G,V]$, from which
pressure and entropy at finite temperature can be obtained \cite{vp1}.

We performed calculations for the internal energy per particle in two ways:
from the Galitskii-Koltun's sum rule (\ref{gksumrule}) and from diagram
summation, i.e. eqs. (\ref{eq:kin}) and
(\ref{eq:final}). For the two methods we employ both a realistic 
nucleon-nucleon interaction, the CD-Bonn potential, and the effective
interaction $V_{low-k}$ from \cite{bodemu}.
\begin{figure}[h]
\begin{center}
\includegraphics[height=7cm]{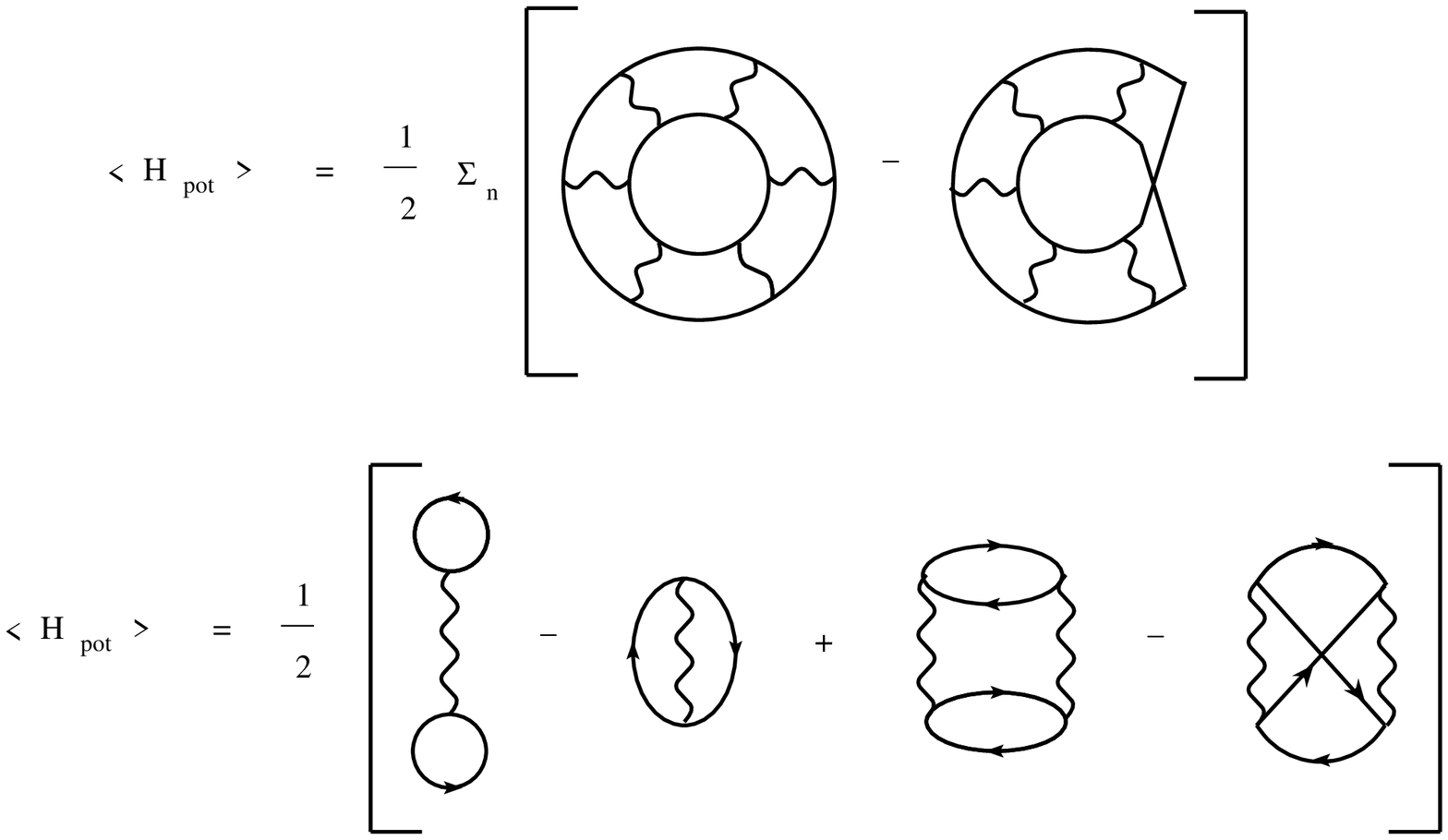}
\caption{Diagrammatic expression for the expectation value 
of the  potential energy in the case of the full $T$-matrix approach
(upper figure, 
the sum runs over the number of interaction lines $n$ in the diagram)
and for a second order approximation (lower figure).}
\label{fig:hpot}
\end{center}
\end{figure}
In the first case we consider the whole expansion of the propagator in
ladder-type diagrams, which lead to the diagrammatic expression for the 
interaction energy that appears in Fig. \ref{fig:hpot} (up). 
The energy depends on dressed propagators (solid lines in the figure) and
on the two-body potential (wavy lines).
In the case of the effective interaction we only 
include diagrams up to the second order (Born approximation), which are
shown in Fig. \ref{fig:hpot} (down). 
Since $V_{low-k}$ already accounts for the correlations in the dense system
the higher order diagrams do not yield significant contributions
and can be neglected \cite{bodemu}.

\begin{figure}[h]
\begin{center}
\includegraphics[width=8cm]{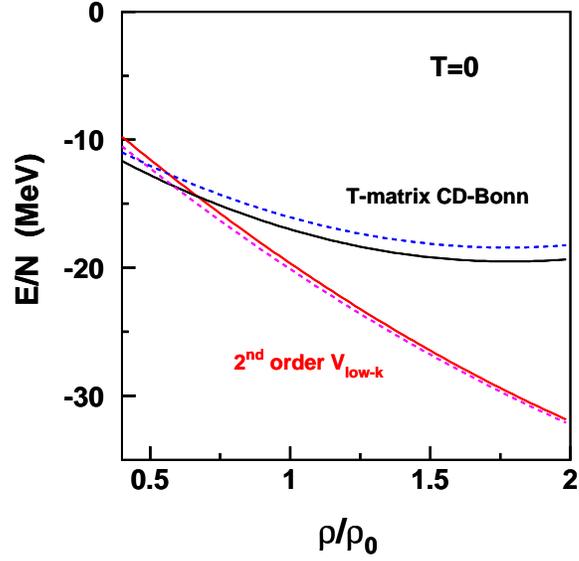}
\caption{The internal energy per particle at zero temperature 
as a function of the density (in units of the empirical
saturation density $\rho_0=0.16$ fm$^{-3}$). The solid
lines represent the expectation value of the Hamiltonian 
(\ref{eq:kin})+(\ref{eq:final}), the dashed lines are the results obtained 
from the sum rule (\ref{gksumrule}).}
\label{fig:eden}
\end{center}
\end{figure}

We restrict ourselves to symmetric nuclear matter at zero temperature,
though we remark that this formalism can be applied to finite temperature
systems as well.
Results for the internal energy per particle as a function of density
are shown in Fig. \ref{fig:eden}, in which calculations from the 
diagram summation are 
compared to the Galitskii-Koltun's sum rule. 
The calculations for the two potentials are in agreement up to $0.8 \: \rho_0$. 
At higher densities, the low-momentum calculations give too attractive 
energies and differ substantially from the full $T$-matrix ones, exhibiting
no minimum. On the contrary, the full calculation with a realistic 
nucleon-nucleon potential shows a saturation point, which however does not
coincide with the experimental value
(as expected since three-body forces are not included.)
\\

Concluding, we have studied the equation of state of symmetric
nuclear matter both with the two-body CD-Bonn potential and with a
renormalized interaction $V_{low-k}$.
When the effective potential is used, the energy behavior shows that
$V_{low-k}$ alone at high densities does not correctly account for all
correlations. This would not happen if a larger cutoff were chosen, even if
in this case $V_{low-k}$ would depend on the starting nucleon-nucleon 
realistic force. This dependence, which can be amplified or suppressed by 
changing the cutoff, can possibly help in testing the reliability of the 
different models of the bare nucleon-nucleon interaction.

\end{document}